# Experimental realization of photonic topological insulator in a uniaxial metacrystal waveguide

Wen-Jie Chen[1,2], Shao-Ji Jiang[1], Xiao-Dong Chen[1], Jian-Wen Dong[1,3,*], C. T. Chan[2]

1. State Key Laboratory of Optoelectronic Materials and Technologies, Sun Yat-Sen University, Guangzhou, China

2. Department of Physics and the Institute for Advanced Study, The Hong Kong University of Science and Technology, Hong Kong, China

3. School of Physics and Engineering, Sun Yat-Sen University, Guangzhou, China

* Email: dongjwen@mail.sysu.edu.cn

Achieving robust transport has attracted much attention in both electronic and photonic systems. Time-reversal invariant photonic crystal can be employed to realize backscattering-immune chiral state, but the chiral bandgap is topologically trivial. Here, we designed and fabricated a metacrystal comprising non-resonant meta-atoms sandwiched between two metallic plates and the system exhibits topological phase transition as established by group classification. The nontrivial bandgap was confirmed by experimentally measured transmission spectra, calculated nonzero spin Chern number and $Z_2$ index. Gapless spin-filtered edge states were demonstrated experimentally by measuring the magnitude and phase of the fields. The transpsort robustness of the edge states were also observed when an obstacle is introduced near the edge.

A signature of topological states [1-6] is the presence of topologically protected edge states, resulting in a quantized conductance. Intense efforts have been devoted to realize similar robust states in photonic systems. For example, a chirality based robust transport was first observed in time-reversal (TR) invariant photonic crystals [7] that are topologically "trivial". Nontrivial bandgap characterized by nonzero Chern numbers has been achieved in TR broken photonic systems with either external magnetic field [8-10] or dynamical modulation [11]. Coupled helical waveguide array [12] was recently proposed to explore topological state by breaking z-reversal symmetry.

TR invariant topological state of quantum spin Hall insulator has been mapped to photonic systems, such as coupled resonator optical waveguides [13-15] and bianisotropic metacrystals [16]. Either clockwise/anticlockwise modes or in-phase/out-of-phase transverse electric (TE) - transverse magnetic (TM) mixed waves are a TR pair of decoupled states, regarded as pseudo spin-up/spin-down states. By either tuning the coupling phases between resonators or introducing bianisotropic effect, the corresponding Hamiltonians take the same form as that of quantum spin Hall effect. In principle, bulk bands for two pseudo spins can acquire nonzero Chern numbers with opposite signs, allowing for the emergence of spin-filtered one-way edge states.

However, the experimental observation of photonic topological insulator (PTI) in the previously suggested photonic metacrystal is challenging, because bianisotropic effect in resonant metamaterials [17, 18] is typically tiny. The ε/μ-matching condition is also not realistic in practice due to high dispersion and loss near the resonant frequency. In this work, we experimentally realize a PTI in a uniaxial metacrystal waveguide -- a hexagonal lattice with non-resonant and non-bianisotropic meta-atoms bounded by two metallic slabs. The field inhomogeneities of the first order waveguide modes along the z direction (perpendicular to the slabs) will naturally introduce a cross coupling between TE and TM modes, which acts as an "effective" bianisotropic effect. The non-resonant meta-atoms are broadband, mildly dispersive, and lossless materials [19-21]. We observed topologically nontrivial bandgaps in uniaxial metacrystal waveguides, which cannot occur in 2D anisotropic photonic crystals. The relative bandwidth is large enough to ensure the existence of spin-filtered edge state and its robust transport behavior. Measured phase differences (between $E_z$ and $H_z$ fields) for the rightward edge state is a constant in the nontrivial bandgap, while the one for the leftward edge state is another constant. The difference between two constants is 180°, manifesting the spin-filtered feature of gapless edge states.

To illustrate the effective bianisotropic effect in the uniaxial waveguide, we consider a spin-degenerate uniaxial medium with $\bar{\bar{\varepsilon}}_r = \rho\bar{\bar{\mu}}_r$ and $\bar{\bar{\mu}}_r = diag\{\mu_{//}, \mu_{//}, \mu_z\}$ sandwiched by two perfect electric conductor (PEC) plates at $z=0$ and $z=d$. We assume that the permittivity and permeability tensors are the functions of (x,y), but invariant in the z direction. The ratio $\rho = [\varepsilon_r]_{ij}/[\mu_r]_{ij}$ is a constant in the whole space. Then, the Maxwell equations have the form of:

$$\nabla \times \vec{E} = i\omega\mu_0\vec{\mu}_r\vec{H}$$
$$\nabla \times \vec{H} = -i\omega\varepsilon_0\rho\vec{\mu}_r\vec{E}$$
(1)

For the m-th order modes in PEC waveguide, we have:

$$\vec{E} = [e_x \sin(\frac{m\pi}{d}z), e_y \sin(\frac{m\pi}{d}z), -e_z \cos(\frac{m\pi}{d}z)]^T$$

$$\vec{H} = [-h_x \sin(\frac{m\pi}{d}z), -h_y \cos(\frac{m\pi}{d}z), h_z \sin(\frac{m\pi}{d}z)]^T,$$
(2)

where $e_x$, $e_y$, $e_z$, $h_x$, $h_y$, and $h_z$ are the functions of (x,y) only. For compactness, we write $\vec{e} = (e_x, e_y, e_z)^T$ and $\vec{h} = (h_x, h_y, h_z)^T$. This form automatically satisfies the PEC boundary conditions. Equation (1) can be rewritten:

$$\nabla \times \vec{e} = i\omega[\mu_0\vec{\mu}_r\vec{h} + \vec{\xi}_e\vec{e}],$$

$$\nabla \times \vec{h} = -i\omega[\varepsilon_0\rho\vec{\mu}_r\vec{e} + \vec{\xi}_e\vec{h}].$$
(3)

where $\vec{\xi}_e$ is an effective magnetoelectric coefficient tensor, with the non-zero elements of $\xi_{e,12} = \xi_{e,21}^* = im\pi/\omega d$ for $m \neq 0$. Denoting a vector $\vec{p}^\pm \equiv \sqrt{\rho\varepsilon_0}\vec{e} \pm \sqrt{\mu_0}\vec{h}$, we can establish a pair of two decoupled spin-polarized states $(p_x^-, p_y^-, p_z^+)$ (spin-up) and $(p_x^+, p_y^+, p_z^-)$ (spin-down) which are linked by TR symmetry. Equation (3) can be rewritten in terms of $\psi^\pm \equiv p_z^\pm / \sqrt{\gamma}$,

$$\{-[\nabla + i\tilde{A}^\pm(\vec{r})]^2 + \tilde{V}(\vec{r})\}\psi^\pm = 0.$$
(4)

These are exactly the two equations for wave functions of a nonrelativistic particle in opposite vector potentials $\tilde{A}^\pm(r)$ but same scalar potentials $\tilde{V}(\vec{r})$ [22]. So a PEC waveguide filled with spin-degenerate uniaxial medium can be mapped to a photonic quantum spin Hall system, when the non-zero order mode is considered. The permittivity and permeability tensors are spatially dependent, and in practical realization one may construct a PTI using a periodic array of non-bianisotropic cylinders. Although the effective magnetoelectric tensor $\xi_e$ is frequency-dependent, it will not change the topological character because it is determined only by the topological invariants as long as the ε/μ-matching condition is satisfied.

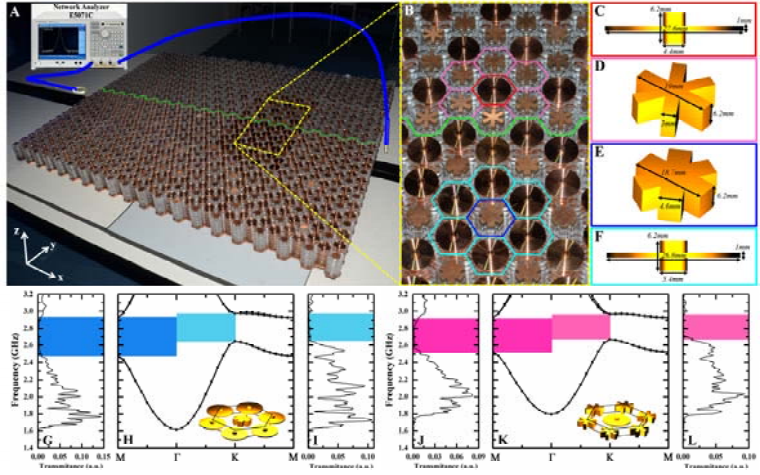

**Fig. 1.** Photonic topological insulator (PTI) and photonic ordinary insulator (POI) are waveguides filled with hexagonal metacrystals. (A) Photograph of two experimental samples with the top copper plate removed to show the geometry inside. Spin-filtered EM waves are guided at the edge (green line) between the insulators. (B) Zoom-in near the edge. The hexagonal unit cells of PTI (blue and cyan hexagons) and POI (pink and red hexagons) are composed of "gyro" and "star"-shaped non-resonant meta-atoms. Both cells are repeated six times in the z direction. Black frames in the insets of (H) and (K) outline the single-layer cell. (C) and (D) are geometry dimensions for POI, while (E) and (F) are for PTI. (G)-(I) Measured transmission spectra and calculated band structure of the PTI are consistent with each other along the ΓM and ΓK directions. (J)-(L) Results for POI.

Let us consider a hexagonal metacrystal waveguide with a three-dimensional unit cell (inset of Fig. 2D). The rod (solid blue) is surrounded by the background material (translucent cyan). The rod has a hexagonal profile with the size of $a$. The lattice constant and the height of waveguide are $\sqrt{3}a$ and $1.6a$, respectively. Figure 2A is the band structure for the first order modes of a topologically nontrivial metacrystal waveguide. All the bands are doubly spin-degenerate for the reason of ε/μ-matching material and inversion symmetry. A complete bandgap exists from 0.264 to 0.293 ($c/a$) (blue in Fig. 2A). Although zero order modes are also allowed in the interested frequency range, they will not couple to other order modes because of orthogonality.

We calculate the $Z_2$ index $\nu_0$ [23] by analyzing the parities of the lowest spin-up band at four Kramers points. Figure 2, B and C, shows the even and odd parities of $E_z$ fields at M and Γ points that are marked by red and blue circles in Fig. 2A. As there is only one odd spin-up mode below the complete bandgap, the bandgap is nontrivial and has a nonzero $Z_2$ index ($\nu_0 = 1$). But the bandgap can be either topologically nontrivial or trivial, depending on the constitutive parameters of anisotropic media. Figure 2D is a phase diagram of the bandgap. The horizontal axis is the ratio of the x-component of permeability between rod and background, while the vertical axis is the ratio of the z-component. The gray area, representing the absence of a complete bandgap, partitions the phase diagram into three regions with trivial/nontrivial bandgap. Figure 2E shows the band structure with the complete bandgap marked with magenta color for a configuration corresponding to a photonic ordinary insulator (POI), which has the identical geometry as the PTI except the constitutive parameters are taken from the purple cross in Fig. 2D. The POI has a trivial $Z_2$ index ($v_0 = 0$) as all the spin-up modes at Kramers points are odd. It possesses a trivial phase and the gapped edge states will appear [22].

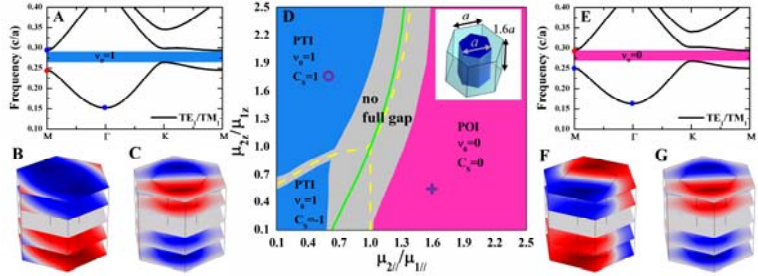

**Fig. 2.** Topological phase transition and nontrivial/trivial bandgaps in the conceptual metacrystal waveguide design. A hexagonal lattice of anisotropic rods is embedded in a PEC waveguide. The field inhomogeneities give rise to the cross coupling between $TE_1$ and $TM_1$ modes and open a topologically nontrivial bandgap. (A) Band structure of PTI. The constitutive parameters are from the purple open circle in (D) with the value of $\varepsilon_2 = 13\mu_2$, $\mu_2 = diag\{0.39, 0.39, 0.44\}$ in the rod and $\varepsilon_1 = 13\mu_1$, $\mu_1 = diag\{0.67, 0.67, 0.25\}$ in the background. (B)/(C) $E_z$ eigenfield at M/Γ point on five z-planes for the lowest spin-up band. The even symmetry at M point ensures the nonzero spin Chern number and $Z_2$ index, leading to the nontrivial bandgap. (D) Phase diagram for the lowest bandgap. Blue and red regions indicate the nontrivial and trivial complete gap, and gray region represents the partial gap. The solid and dash curves show the mode exchange at M and K points. The transition between PTI and POI coincides with a gap reopening and mode exchange at M point, due to the different topologies of two insulators. (E) Trivial band structure of POI. The constitutive parameters are from the purple cross in (D) with the value of $\varepsilon_2 = 13\mu_2$, $\mu_2 = diag\{0.72, 0.72, 0.22\}$, $\varepsilon_1 = 13\mu_1$, and $\mu_1 = diag\{0.45, 0.45, 0.41\}$. (F)/(G) Odd modes at Kramers points in the POI.

Because of the topological nature, each phase in the diagram is an isolated "island". The nontrivial phase cannot be adiabatically connected to a trivial phase without closing and reopening the bandgap. Green solid line in Fig. 2D highlights the band inversion at M point. Two pairs of modes at M point with different parities exchange their positions, involving the bandgap's closing and reopening. The $Z_2$ index changes together with the topological phase. Furthermore, when $\mu_{2//}/\mu_{1//}<1$, there are two PTI phases due to the band inversion at K point (yellow dashed line in Fig. 2D). This exchange of modes will not change the topology as $v_0=1$ is maintained, but it changes the spin Chern number from 1 to -1. The numbers of gapless edge states in both PTI phases are one, but the propagation directions reverse.

To experimentally realize the metacrystal waveguide PTI, we constructed samples using non-resonant meta-atoms as shown in Fig. 1A. The meta-atoms have the same "gyro" and "star" geometries, but are arranged in conjugate patterns for POI (upper half in Fig. 1B) and PTI (lower half in Fig. 1B) and different geometry dimensions (Fig. 1). With properly designed parameters and meta-atom arrangements [22], all PTI bands are almost doubly degenerate from 1.6 to 3.2 GHz (Fig. 1H), indicating that the ε/μ-matching condition is met for a broad bandwidth. A nontrivial bandgap spanning from 2.65 to 2.93 GHz is achieved for $v_0=1$. Microwave transmission spectra in Fig. 1, G and I, confirmed theory-predicted bandgaps along ΓM and ΓK directions. For the POI (Fig. 1K), the photonic bands are almost identical to those of PTI. However, the POI bandgap is trivial and the topological classification is different from the PTI.

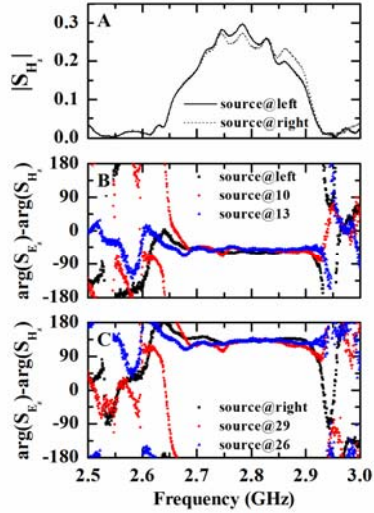

**Fig. 3.** Propagation characteristics of the spin-filtered states at the edge between PTI and POI. (A) Measured transmission for the rightward (solid) and leftward (dotted) gapless topological states. (B) Phase differences between the $E_z$ and $H_z$ components at the right exit when the source is placed at either the left entrance (black), in the 10[th] hole (red), or in the 13[th] hole (blue). An obvious phase plateau at -50° appears in the complete bandgap no matter where the source locates, since the spin-up rightward state is the only exited edge state. (C) Same as (B) except that the exit is on the left and the source is placed at the right entrance (black), in the 29[th] hole (red), or in the 26[th] hole (blue). The phase differences stabilize around 130°. Both (B) and (C) demonstrate the spin-filtered feature of the gapless topological states.

Spin-filtered gapless edge state is a signature feature at the boundary of two materials with different topological phases. The edge is realized by placing two metacrystal waveguides side-by-side (green line in Fig. 1A). To excite the rightward state, an $H_z$-polarized source through a loop antenna was placed at the left. The $E_z$ and $H_z$ transmitted wave are measured at the right end, by a monopole and a loop antenna, respectively (see SOM). High transmission for the rightward state, where the frequency range lies within the overlapped bandgap of both insulators, demonstrate the gapless property (Fig. 3A). Moreover, the edge is spin-filtered. When the source is at the left, only a spin-up state can be excited, meaning that the $E_z$ and $H_z$ fields are in phase throughout the whole edge [24]. Due to impedance mismatch at the exit, additional phases will mix into the signal [25]. The measured phase difference between the $E_z$ and $H_z$ fields of the gapless topological state will keep constant but deviate from 0°. This is verified by the phase plateau around -50° from 2.68 to 2.92 GHz (black in Fig. 3B). In addition, as the spin-up state is the only rightward mode, the phase difference at the right exit should not vary no matter where the source is. This is also demonstrated by the red and blue curves in Fig. 3B (see SOM). For the leftward spin-down state, it can be excited if swapping the positions of source and detector. High transmission is still present due to time reversal symmetry (dotted line in Fig. 3A). The phase differences at the left exit are stable around 130° independent of source position (Fig. 3C).

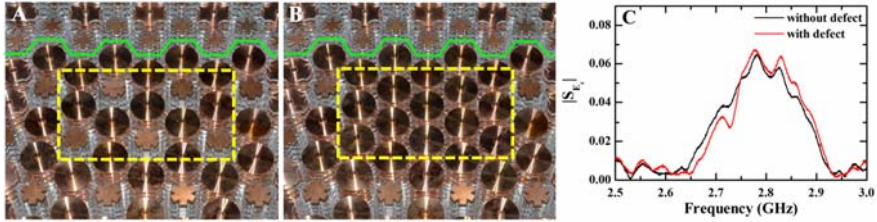

**Fig. 4.** Robust transport at the edge of PTI. (A)/(B) Close-up view of the sample with/without defect. (C) Transmission for the edge between PTI and POI. Black is for the case of no defect, while red is for the defect with five unit cells of star meta-atoms being substituted by gyro meta-atoms. No obvious backscattering is introduced by the defect. The right-going spin-up edge state keeps moving rightward even when it encounters a defect, due to the absence of left-going spin-up edge state.

To demonstrate transport robustness at the edge between two metacrystal waveguides, we measured the transmission with/without disorder. The disorder is introduced by substituting five unit cells of star meta-atoms by gyro meta-atoms near the center of the edge (dashed panes in Fig. 4). Figure 4C illustrates that high transmission is maintained when the disorder is present. No backscattering is introduced by the defect for the spin-filtered feature. When the right-going spin-up state encounters a disorder, it has no choice but to keep moving rightward for the absence of left-going spin-up state.